# Scaling up archival text analysis with the blockmodeling of n-gram networks: A case study of Bulgaria's representation in the *Osservatore Romano* (January – May 1877)

✉ Fabio Ashtar Telarico

ORCID ID: 0000-0002-8740-7078
University of Ljubljana, Faculty of Social Sciences
Kardeljeva ploščad 5
Ljubljana, Slovenia
E-mail: Fabio-Ashtar.Telarico@fdv.uni-lj.si

**Abstract**

This paper seeks to bridge the gap between archival text analysis and network analysis by applying network clustering methods to analyze the coverage of Bulgaria in 123 issues of the newspaper Osservatore Romano published between January and May 1877. Utilizing optical character recognition and generalized homogeneity blockmodeling, the study constructs networks of relevant keywords. Those including the sets Bulgaria and Russia are rather isomorphic and they largely overlap with those for Germany, Britain, and War. In structural terms, the blockmodel of the two networks exhibits a clear core-semiperiphery-periphery structure that reflects relations between concepts in the newpaper's coverage. The newspaper's lexical choices effectively delegitimised the Bulgarian national revival, highlighting the influence of the Holy See on the newspaper's editorial line.

**Keywords:** network analysis, blockmodeling, Russia, Bulgaria, text analysis

## 1. Introduction

In an increasingly data-driven world, archival work challenges researchers to handle growing amounts of data from diverse sources and yet keep producing deep qualitative analysis. And this challenge is particularly pressing in dealing with inherently complex textual data consisting of thousands or even millions of interconnected terms, concepts, and ideas. Understanding these complex relationships is crucial for making sense of the vast amounts of information generated daily.

Against this background, network analysis (NA) provides a powerful framework for handling large corpora at several, complementary levels of analysis. By representing words and/or documents as nodes bound together in a network by some predeterminable sort of tying relationship, NA offers tools to explore and analyse the structure of interconnectedness within texts and between documents. By granting the ability to analyse texts in a fraction of the time required to read them, NA can yield systematic analyses of the semantic relationships between words in large corpora beyond the limitations of traditional qualitative approaches.

However, this approach is not widely used in archival research despite the ever-growing number of archives opening their doors to the public. In fact, most analyses of early- and pre-20[th] century corpora delve into historical disquisitions over linguistics, societal norms, and specific intellectual movements or even individuals using predominantly qualitative methods. And this is especially true for the state of the art in Bulgaria. Partly, this lack of interest into NA and other computational methods stemmed from the difficulty of digitalising old texts starting from the low-quality scans researchers have access to or can afford to acquire. But new optical-character recognition technologies can achieve satisfactory levels of reliability, thus invalidating once-justified concerns. Rather, the main obstacle remains the lack of cross-fertilisation between NA and historical-text analysis. In fact, NA can have significant downstream applications in text analysis of which many archival researchers are unaware.

In order to help foster a healthy dialogue between the two fields, this paper explores the use of network-clustering methods for text analysis through a case study of 123 issues of the newspaper *Osservatore Romano* published between January and May 1877. Namely, this study attempts to shed a light on the existence and, conditional on that, meaning of the newspaper's coverage of Bulgaria over the period under consideration. The choice of this particular publication is informed by both data-availability and the prominent role that the Kingdom of Italy and the Holy See, which financed the newspaper since 1871, had in shaping and framing the so-called 'Eastern Question'. Moreover, the year 1877 is of pivotal importance in Bulgarian history as it marked the beginning of the fifth Russo-Ottoman war culminating in Congress of Berlin and the independence of a shrunk Principality of Bulgaria. The paper leverages this empirical application to test the use of computational and NA methods on older archival sources that pose additional issues such as low-quality inputs or archaic variations in spelling, grammar and writing style. Moreover, it answers the research question showcasing the use of a network-clustering technique called blockmodeling which is particularly apt to uncover groupings of words and sematic patterns amongst them thereby allowing a qualitative understanding of large textual datasets.

The substantive and methodological results are of immediate relevance for the field. On the substantive plan, it is shown that the *Osservatore Romano*'s coverage of Bulgaria echoes that of other Catholic opinion-making institutions around Europe. In particular, it highlights militant Catholics' attempt at delegitimising the independence struggles in the Balkan provinces of the Ottoman Empire through a combination of under-reporting, politically laden lexical choices, and focus-shifting. In methodological terms, the results paper highlights that NA is a key tool in knowledge extraction from large corpora and that blockmodeling can produce a nuanced understanding of textual data ultimately facilitating data-driven strategies across disciplines.

## 2. Methods

### 2.1 Text pre-processing: Feeding a 19th century newspaper to a 21st century algorithm

Computational text analysis begins with text pre-processing is an essential step for applying computer tools to analogical textual data.

In the case at hand the first issue was the relatively low quality of the archival scans relative to the broadsheet size of each issue's four pages (approx. 60x75cm). Most OCR software, including some paid ones, has very low reliability on this type of images, especially given that the print degraded over the centuries. Instead, the open-source OCR Kraken (see Kiessling 2019) provides binarization, segmentation, box-bounding and recognition for old texts and is optimised to handle imperfections of this type.[i] By manually testing the first eight pages (two issues, approx. two percent of all words), Kraken's accuracy nears 98% if words misspelled due to the deterioration of the print (e.g., an 'n' looking like an 'r' because the second leg faded away) are counted as correct or 90% otherwise.

Once a text file is obtained, it has to be cleaned of noise by: spellchecking, case normalisation (typically to lowercase), and removing stopwords (common words like articles, prepositions, and auxiliary verbs), special/typographic characters, and punctuation marks (HaCohen-Kerner et al. 2020 p. 9ff). Spellchecking was executed using a novel probabilistic spellchecker[ii] trained on a corpus of over 16 million words called *ChroniclItaly* (Viola & Fiscarelli 2021). The spellchecker (tested on approx. two percent of all text) corrected almost all of the 'old-age' errors made by the OCR and all but two of the one-character errors. Overall, final-text accuracy is around 97%. Text cleaning was executed using the package morestopwords for R (Telarico & Watanabe 2023). More advanced, but equally necessary steps include word stemming or lemmatisation. Linguistically, lemmatisation[iii] 'is the task of grouping together word forms that belong to the same inflectional morphological paradigm and assigning to each paradigm its corresponding canonical form called lemma.' (Gesmundo & Samardžić 2012 p. 368)

## 2.2 Enter n-grams: From lemmas to networks

Preprocessed text is typically represented numerically, with one popular method being the Bag of Words (BoW) model (Zhang et al. 2010). A BoW is a (generally sparse) matrix where each column stands for a document and each row for a word. The cells indicate whether (or how many times) a lemma is present in each document. Clearly, this matrix represents a two-mode networks of documents and texts which poses peculiar analytical problems well-discussed in the NA literature.

Thus, the task can be simplified by switching to the n-gram model (Pereira et al. 1999), which captures the probabilistic relationship between lemmas by grouping them based on their co-occurrence. The matrix is usually less sparse than a BoW and, in the case of bi-gram, represents a one-mode network of pairs of words where the tie is defined as following/being followed by another word with a weight equal to the count of co-occurrences or its normalised probability. It is especially sensible to use bi-grams rather than the BoW because the focus is here on a relatively small part of the text (that connected to Bulgaria) and there are only 123 documents. The result is a typical hairball with 'significant node occlusion and [many] link crossings' (Edge et al. 2018 p. 3951).

Practically, the bi-gram network so obtained has to be reduced in size by cutting out the second-order neighbourhoods of the keywords in Table 1. Sub-networks of more than 500 lemmas/nodes were reduced by degree centrality (only lemmas with at least as many connections as the median of the lemmas with more than two neighbours). If still larger than 500 units, the networks were shrunk further by k-core decomposition (the smallest core with no less than 95 nodes).

| Set | Keyword | Examples | Cont. | Keyword | Examples |
|---|---|---|---|---|---|
| Bulgaria | ^bulg.* | Bulgaria, bulgaro | Germany | ^bismark* | Bismark, bismarkiano |
| Balkans | ^balcan.* | Balcani, balcanico | | ^prussi* | prussia, prussiano |
| Slav | ^slav.* | Slavo, slavica | | ^tedesc.* | tedesco, tedeschi |
| Turkey | ^ottoman.* | Ottomani, ottomano | Britain | ^ingles* | inglese, inglesi |
| | ^turc* | Tuchia, turco | | ^britan* | britannico |
| | ^turchi* | Tuchia, turchi | | ^londra | londra |
| Russia | ^russ.*$ | Russia, russo | War | ^guerr* | guerra, guerrafondaio |
| | ^zar.* | zar, zarina | | ^bellic* | bellico, bellicosa |
| | ^romanov | romanov | | | |

Table 1 List of keywords looked-up in the text and their aggregation in sets.

## 2.3 Generalised Blockmodeling: Looking for patterns and groups

In NA clustering is a collection of methods for classifying nodes into small, interpretable groups. Formal clustering of a bi-gram network yields information of the pattern of relations between, making it easier to understand and analyse the content, structure, and themes of all underlying texts. This helps uncover meaningful relationships within the network, leading to a deeper understanding of how language is used.

Generalised blockmodeling is an approach to clustering networks designed to systematically identify and model 'clusters of equivalent units based on a selected definition of equivalence.' (Žiberna 2007 p. 105) While it has not found commonplace application in the field, the generalised blockmodeling of valued and weighted networks (developed by Žiberna 2007) is made accessible to researchers from all disciplines by the graphical interface `BlockmodelingGUI` (Telarico & Žiberna 2022). And this method is particularly useful for analysing text data as it can identify clusters (or 'blocks') exhibiting distinct relational patterns. Unlike other types of blockmodeling, generalised blockmodeling requires the researcher to feed both the desired number of clusters and the connections allowed between them (block types)[iv] to the algorithm. Therefore, an in-depth understanding of the data (in this case, the texts) and theoretical understanding of the expected result guide the definition of these parameters. Then, an objective function evaluates the quality of the clustering measuring how well the result matches the desired intra- and inter-block structure. The parameters can then be updated dialectically as intermediate results are produced. The bi-grams network around the chosen keyword-sets was modelled using so-called 'homogeneity' blockmodeling, which 'searches for the partition where the sum of some measure of within block variability over all blocks is minimal.' (Žiberna 2007 p. 105)

## 3. Results

All networks were blockmodeled allowing null and complete blocks only. But those built around the keywords-set (B) Bulgaria, (R) Russia, Germany, and Britain were optimised for six clusters and the others for three due to a large difference in size. After optimising the partitions, the final results for the keyword-sets *Russia* and *Bulgaria* are rather isomorphic (cf. Figure 1), and they largely overlap with those for Germany, Britain, and War.[v]

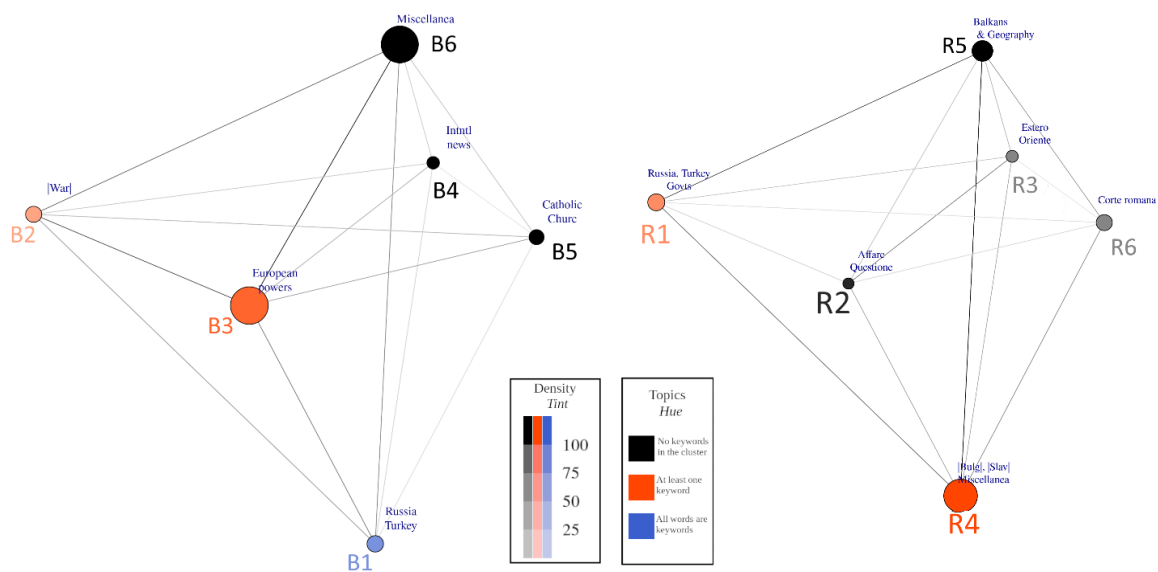

Figure 1 Blockmodel of the reduced bi-gram network built around the keyword-set Bulgaria and Russia with labelled clusters.

In terms of network structure, network B exhibit a clear core-semiperiphery-periphery structure (see Figure 1) with the lemmas Russia-Turkey at the centre (cluster B1), War the first semiperiphery (B2), and words like 'peoples', '(to) declare', 'power', and 'minister' in the second semiperiphery (B3). The clusters made up of 'News' plus 'International' (B4) makes up a bridging core connected to B1. Whereas B2 is a semiperiphery shared between B1 and the core comprising 'Catholic' and 'Church' (B5).

Similarly, the network R showcases a similar structure articulate around the core cluster 'Russia' plus 'Turkey' plus 'government' (R1) with two peripheries made up of 'Affairs' plus 'Question' (R2) and 'International' plus 'Eastern' (R3). A sparser periphery including 'Bulgaria' and 'Slav' as well 'destiny', 'protest', 'kingdom', 'revolution', 'repel', and 'march' occupies cluster R4. Whereas cluster R5 includes the keyword 'Balkans' plus some political and geographical/naturalistic terms.

## 4. Discussion

The analysis shows that the reporting was largely reactive to recent events, yet it consistently emphasized the dynamics of great-power politics. The primary reference points in the analysis were the major powers of the time, particularly Germany, Britain, and Russia – which sits in the core of the blockmodel. Moreover, the high degree of isomorphism between the B and R partitions, together with the vast overlapping with the networks built around Germany and Britain highlights the fact that even by trying to focus on Bulgaria, one cannot avoid to pass through other international issues. Namely, the newspaper focused predominantly on the great-power competition between Germany, Russia, and Britain. And even the Ottoman Empire does nor get many mentions if not in relation to its rival, reflecting the broader geopolitical tensions of the era.

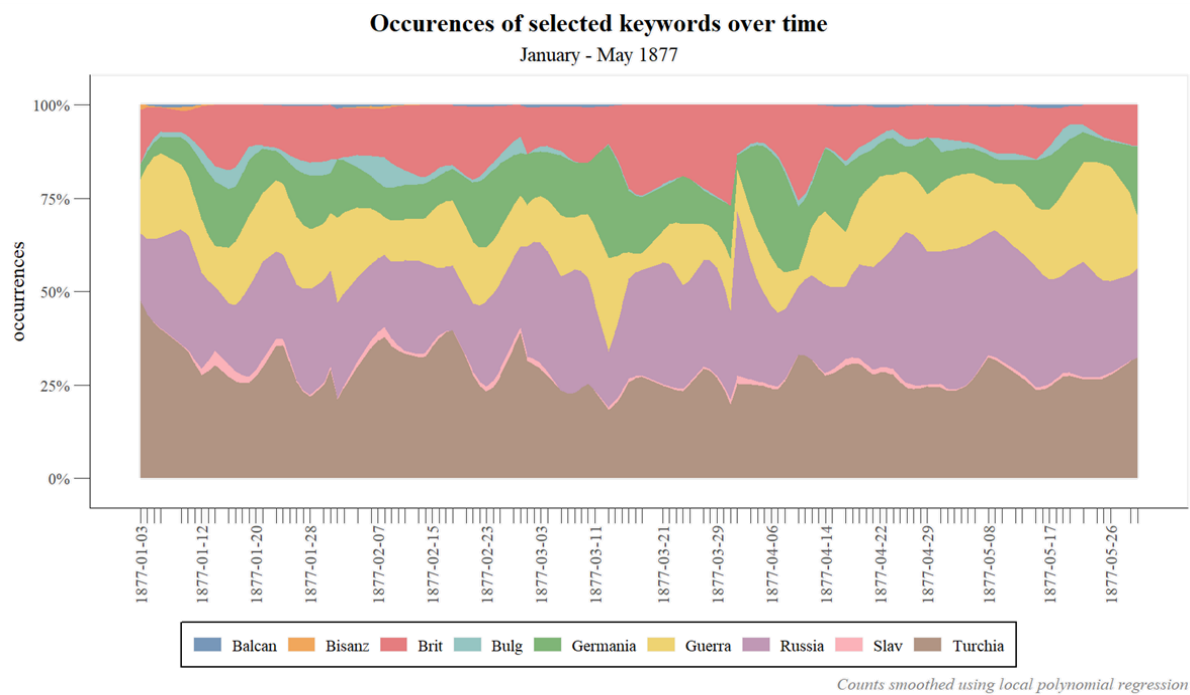

Figure 2 Relative occurrence of the keyword-sets in the corpus over time.

Looking at the occurrence of keyword over time (Figure 2), the newspaper's concern with great-power competition and, especially, Russia's hostile intention vis-à-vis the Ottoman Empire is evident. In fact, the use of lemmas belonging to the aforementioned keyword-sets is reactive to events on the ground such as the failure of the Constantinople Conference in January, the Russian declaration of war in April, and the subsequent war operations in the Romanian theatre including the Sinking of the battleship Lütf-ü Celil & the monitor Seyfi.

Additionally, the pattern of ties in both the B and R blockmodels shows that the *Osservatore Romano* systematically tied Bulgaria to other emerging Balkan states (notably Serbia). This suggests that the Bulgarian issue was not addressed on its own merits, but rather treated as a piece on a much greater geopolitical chessboard. The newspaper's lexical choices further reinforce this trend. The use of words like 'province' instead of 'nation' and 'revolution' over 'liberation' was instrumental in delegitimising the emancipatory significance of the Bulgarian independence struggle. The omission of perspectives other than those of the great powers – which, except for Russia, supported the Ottoman Empire and the continued oppression of aspiring independent nations in Eastern Europe – highlights the influence of the Vatican over the newspaper's editorial line. And this finding is in line with observations made in studies addressing other countries with more traditional methods (Rossi 1982 p. 55).

This case study demonstrates that archival researchers can leverage computational methods even when working with pre-20th century texts, utilizing tools like Kraken OCR. The ability to convert historical texts into digital form, coupled with the computational power to study them through network analysis, opens up new avenues for scholarly investigation. This expands the scope of traditional textual analysis, providing deeper insights into historical documents.

The extensive literature on network analysis applied to texts, along with accessible software like `BlockmodelingGUI`, further highlights the potential for network analysis to become part of the text analysts' repertoire. Moreover, incorporating direct blockmodeling of BoW and n-gram networks into text analysis represents a continuation of the trend toward the consolidation of other network-analytical methods such as topic modelling. This integration enables researchers to explore the semantic, structural, and relational aspects of texts more comprehensively, enhancing the depth and breadth of textual analysis. Nevertheless, further work is necessary to expand the corpus longitudinally and latitudinally, as well as improve the text pre-processing methods' accuracy.

---

[i] Since Kraken provides bounding boxes with a rectangular shape, it is helpful to minimise segmentation errors by and de-skewing and de-warping the images.

[ii] The spellchecker develops the idea proposed by Peter Norvig (2016), director of research at Google, with some ad-hoc adjustments.

[iii] This is becoming the standard procedure for most languages, but stemming was prevalent in the past and it is still widely used for inflectional languages.

[iv] There are tens of block types. The most used are: complete blocks, where all nodes within a block are interconnected; regular blocks, where nodes within a block exhibit similar connection patterns; and null blocks, where there are no connections between nodes within a block. The type of block structure identified depends on the network's overall topology and the relationships between its nodes.

[v] The clusters in the Turkey, Balkans, Slav, and War networks are more heterogenous and harder interpret.